\documentclass[12pt]{article}

\newcommand{\be}{\begin{equation}}
\newcommand{\ee}{\end{equation}}
\newcommand{\bi}[1]{\bibitem{#1}}
\usepackage{epsfig,amsmath,amssymb,graphics,graphicx} 

\begin{document}
\begin{center}
Communications in Applied Analysis 12 (2008) 441–450
\vskip 5mm

{\Large \bf Fractional Powers of Derivatives in 
Classical Mechanics} 
\vskip 5mm

{\large \bf Vasily E. Tarasov}\\

\vskip 3mm

{\it Skobeltsyn Institute of Nuclear Physics, \\
Moscow State University, Moscow 119991, Russia } \\
{E-mail: tarasov@theory.sinp.msu.ru}

\vskip 11 mm

\begin{abstract}
Fractional analysis is applied to describe classical dynamical systems. 
Fractional derivative can be defined as a fractional power of derivative.
The infinitesimal generators $\{H, \ . \ \}$ and
${\cal L}=G(q,p) \partial_q+F(q,p) \partial_p$,
which are used in equations of motion, are derivative operators. 
We consider fractional derivatives on a set of classical observables 
as fractional powers of derivative operators.
As a result, we obtain a fractional generalization of the
equation of motion.  
This fractional equation is exactly solved for the 
simple classical systems.
The suggested fractional equations generalize a notion 
of classical systems to describe dissipative processes.
\end{abstract}

\end{center}


\section{Introduction}

The analysis of non-integer order 
goes back to Leibniz, Liouville, Grunwald, Letnikov and Riemann.
There are many books about fractional calculus and 
fractional differential equations \cite{SKM,Podlubny,KST}.
Derivatives of fractional order, and 
fractional differential equations 
have found many applications in recent studies in physics
(see, for example, \cite{Zaslavsky1,GM,WBG,Zaslavsky2,MS} 
and references therein).

The classical variables, which are also called observables, 
are defined as functions on the phase space.
The dynamical description of system is given by an operator.
The natural description of the motion is in terms of 
the infinitesimal change of the system.
The infinitesimal operator of equation of motion
is defined by some form of derivation of functions.

Fractional derivative can be defined as a fractional power
of derivative (see, for example, \cite{MartSanz,IJM}).
It is known that the infinitesimal generator $\{H, \ . \ \}$ 
and ${\cal L}=G(q,p) \partial_p+F(q,p) \partial_p$,
which are used in the equation of motion, 
are derivations on an algebra of classical observables. 
A derivation of an algebra ${\cal M}$ 
is a linear map ${\cal L}$, which satisfies
${\cal L}(AB)=({\cal L}A)B+ A({\cal L}B)$ 
for all $A,B \in {\cal M}$.
In this paper, we consider a fractional derivative 
as a fractional power of derivative.
As a result, we obtain a fractional generalization 
of the equation of motion.
It allows us to generalize a notion of classical Hamiltonian systems. 
Note that some fractional generalization of 
gradient systems has been suggested in \cite{FracHam1}, 
and a generalization of Hamiltonian systems is 
considered in \cite{FracHam2}.
The suggested fractional equation is exactly solved 
for a free particle, harmonic oscillator and damped oscillator.
A classical system that is presented by fractional equation 
can be considered as a dissipative system.
Fractional derivatives can be used as a possible approach to 
describe an interaction between the system and an environment.
Note that fractional dynamics can be considered with 
low-level fractionality by some generalization of 
method suggested in \cite{TZ2,TofP,TofG}.

In section 2, the fractional power of derivative and 
the fractional equation are suggested.
In section 3, the Cauchy problem for the fractional equation and the
properties of time evolution described by it are considered.
In section 4, the solution for simple examples of the fractional equation 
are derived. \\

\section{Fractional derivative and fractional equations}

Let us consider the classical systems
\be \label{qp}
\frac{d}{dt}q_k=G_k(q,p), \quad  \frac{d}{dt}p_k=F_k(q,p) \quad
(k=1,...,n). \ee
A linear algebra ${\cal M}$ of classical observables is
described by functions $A=A(q,p)$ on the phase space $\mathbb{R}^{2n}$. 
Let ${\cal L}$ be a differential operator on ${\cal M}$ given by
\be \label{calL}
{\cal L}=-\Bigl( G_k(q,p) \frac{\partial}{\partial q_k} + 
F_k(q,p) \frac{\partial}{\partial p_k} \Bigr) . \ee
Here and later we mean the sum on the repeated index $k$ from 1 to n.
The equation of motion for the classical observable has the form
\be \label{Ham1} 
\frac{d}{dt} A_t=- {\cal L} A_t . \ee
Equations (\ref{qp}) are special cases of (\ref{Ham1}).

If the functions $G_k(q,p)$ and $F_k(q,p)$
satisfy the Helmholtz conditions  
\be \label{HC1} \frac{\partial G_{k}}{\partial p_l}-
\frac{\partial G_{l}}{\partial p_k}= 0, \ee
\be \label{HC2} \frac{\partial G_{l}}{\partial q_k}+
\frac{\partial F_{k}}{\partial p_l}=0, \ee
\be \label{HC3} \frac{\partial F_{k}}{\partial q_l}-
\frac{\partial F_{l}}{\partial q_k}= 0, \ee
then the classical system (\ref{Ham1}) is a Hamiltonian system.
In this case, $G_k$ and $F_k$ can be represented in the form
\[ G_k(q,p)=\frac{\partial H}{\partial p_k}, \quad
F_k(q,p)=-\frac{\partial H}{\partial q_k} .\]
If $H=H(q,p)$ is a continuous differentiable function,
then the conditions (\ref{HC1}), (\ref{HC2}) and (\ref{HC3})
are satisfied. 
The equations of motion (\ref{Ham1}) for Hamiltonian system 
can be written in the form
\be \label{fHH} \frac{d}{dt}A_t=-\{H,A_t\} \ee
where $\{ \ , \ \}$ is the Poisson bracket
\[ \{A,B\}= \frac{\partial A}{\partial q_k} \frac{\partial B}{\partial p_k}-
\frac{\partial A}{\partial p_k} \frac{\partial B}{\partial q_k} . \]
The time evolution of the Hamiltonian system is 
induced by the Hamiltonian $H$.

It is interesting to obtain fractional generalizations
of equations (\ref{Ham1}) and (\ref{fHH}).
We will consider here concept of fractional power for ${\cal L}$.
If ${\cal L}$ is a closed linear operator
with an everywhere dense domain $D({\cal L})$,
having a resolvent $R(z,{\cal L})=(zL_I-{\cal L})^{-1}$ 
on the negative half-axis, 
then there exists \cite{Bal,Yosida,Krein} the operator
\be \label{LaA}
-({\cal L})^{\alpha}=\frac{\sin \pi \alpha }{\pi} 
\int^{\infty}_0 dz\, 
z^{\alpha-1} R(-z,{\cal L}) \, {\cal L}
\ee
is defined on $D({\cal L})$ for $0< \alpha <1$. 
The operator $({\cal L})^{\alpha}$ is 
a {\it fractional power of the operator ${\cal L}$}.
It is known that the linear differential operator (\ref{calL})
is a closable operator \cite{Yosida}.
Note that 
\[ ({\cal L})^{\alpha}({\cal L})^{\beta}=
({\cal L})^{\alpha+\beta} \]
for $\alpha, \beta>0$, $\alpha+\beta<1$. 

As a result, we obtain the equation
\be \label{Ham2} 
\frac{d}{dt} A_t=-({\cal L})^{\alpha} A_t , \ee
where $t$  is dimensionless variable.
This is the {\it fractional equation of motion}.
We can define a fractional generalization of equation (\ref{fHH}) by
\be \frac{d}{dt}A_t=-(\{H, \ . \ \})^{\alpha} A_t \ee
Note that $(\{H, \ . \ \})^{\alpha}$
cannot be presented in the form $\{H^{\prime}, \ . \ \}$
with a function $H^{\prime}$.
Therefore, fractional system described by (\ref{Ham2}) with 
${\cal L}=\{H, \ . \ \}$ are not Hamiltonian systems.
The systems will be called the fractional Hamiltonian systems (FHS). 
Usual Hamiltonian systems can be considered as 
a special case of FHS. 
Note that another fractional generalization of 
Hamiltonian systems has been suggested in \cite{FracHam2}.

\section{Solutions of fractional equations of motion}

\subsection{Cauchy problem for fractional equations}

If we consider the Cauchy problem for equation (\ref{Ham1}) 
in which the initial condition is given at the time $t=0$ by $A_0$,
then its solution can be written in the form
\[ A_t=\Phi_t A_0. \]
The operator $\Phi_t$ is called the evolution operator.
It is not hard to prove that the following properties are satisfied:
\[ \Phi_t \Phi_s=\Phi_{t+s}, \quad (t,s >0) , \quad \Phi_0=I , \]
where $IA=A$ for all $A \in {\cal M}$.
As a result, the operators $\Phi_t$ form a semi-group.
Then the operator ${\cal L}$ is called the generating operator,
or infinitesimal generator, of the semi-group $\{\Phi_t, t\ge 0\}$.

Let us consider the Cauchy problem for 
fractional equation (\ref{Ham2}) 
in which the initial condition is given by $A_0$.
Then its solution can be presented \cite{Yosida,Krein} as
\[ A_t(\alpha)=\Phi^{(\alpha)}_t A_0, \]
where the operators $\Phi^{(\alpha)}_t$, $t>0$, 
form a semi-group which will be called the {\it fractional semi-group}.
The operator $({\cal L})^{\alpha}$ is infinitesimal generator of 
the semi-group $\{\Phi^{(\alpha)}_t, t\ge 0\}$ 
that can be presented by
\[ ({\cal L})^{\alpha}=\frac{1}{\Gamma(-\alpha)} 
\int^{\infty}_0 ds \, s^{-\alpha-1} (\Phi^{(\alpha)}_t-I) . \]
This is the Balakrishnan equation \cite{Bal,Yosida}.

\subsection{Properties of fractional evolution operator}

Let us consider some properties of temporal evolution described by
a fractional semi-group $\{\Phi^{(\alpha)}_t, t\ge 0\}$. \\

(1) The operators $\Phi^{(\alpha)}_t$ can be constructed
in terms of the operators $\Phi_t$
by the Bochner-Phillips formula \cite{BP,Phil,Yosida}:
\be \label{BPf2}
\Phi^{(\alpha)}_t=
\int^{\infty}_0 ds f_{\alpha}(t,s) \Phi_s , 
\quad (t>0) . \ee 
Here $f_{\alpha}(t,s)$ is defined by
\be \label{fats}
f_{\alpha}(t,s)=\frac{1}{2\pi i} \int^{a+\i\infty}_{a-i\infty}
dz \, \exp (sz-tz^{\alpha}) ,
\ee
where $a,t>0$, $s \ge 0 $, and $0<\alpha <1$.
The branch of $z^{\alpha}$ is so taken that $Re(z^{\alpha})>0$
for $Re(z)>0$.
This branch is a one-valued function in the $z$-plane 
cut along the negative real axis.
The convergence of this integral is obviously
in virtue of the convergence factor $\exp(-tz^{\alpha})$.
By denoting the path of integration in (\ref{fats})
to the union of two paths $r\, \exp(-i \theta)$, and 
$r\, \exp(+i \theta)$, where $r \in (0,\infty)$, 
and $\pi/2 \le \theta \le \pi$, we can obtain
\[
f_{\alpha}(t,s)=\frac{1}{\pi} \int^{\infty}_0  dr \,
\exp (sr \cos \theta - t r^{\alpha} \cos (\alpha \theta)) \cdot \]
\be \label{freal}
 \cdot \sin (sr \sin \theta - 
t r^{\alpha} \sin (\alpha \theta)+ \theta) . \ee
If we have a solution $A_t$ of equation (\ref{Ham1}),
then formula (\ref{BPf2}) gives the solution
\be 
A_t(\alpha)=\int^{\infty}_0 ds \, f_{\alpha}(t,s) A_s , 
\quad (t>0) \ee
of fractional equation (\ref{Ham2}). 
As a result, we can obtain solution of fractional equation by 
using well-known solutions of usual equations. \\

(2) In classical mechanics, 
the most important is the class of real operators. 
Let $*$ be a complex conjugation.
If $\Phi_t$ is a real operator on ${\cal M}$, then
\[ (\Phi_t A)^{*} = \Phi_t (A^{*}) \] 
for all $A \in D (\Phi_t) \subset {\cal M}$.
A classical observable is a real-valued function. 
If $\Phi_t$ is a real operator
and $A$ is a real-valued function $A^{*}=A$, 
then the function $A_t=\Phi_t A$ is real-valued, i.e., 
$(\Phi_t A)^{*}=\Phi_t A$. 
An operator, which is a map from a set of observables 
into itself, should be real.
All possible dynamics, i.e., 
temporal evolutions of classical observables,
should be described by real operators. 
Therefore the following statement is very important.
If $\Phi_t$ is a real operator, then $\Phi^{(\alpha)}_t$ is real. 
The proof will follows from the Bochner-Phillips formula, 
which gives
\[ (\Phi^{(\alpha)}_t A)^{*} =
\int^{\infty}_0 ds \, f^{*}_{\alpha}(t,s) (\Phi_s A)^{*} , 
\quad (t>0) . \]
Using (\ref{freal}), it is easy to see 
that $f^{*}_{\alpha}(t,s)=f_{\alpha}(t,s)$ 
is a real-valued function.
Then $(\Phi_t A)^{*}=\Phi_t A^{*}$ leads to 
\[ (\Phi_t A)^{*} = \Phi_t (A^{*}) \]
for all $A \in D (\Phi^{(\alpha)}_t) \subset {\cal M}$. \\

(3) Let $\Phi_t$ be a operator on ${\cal M}$.
An adjoint operator of $\Phi_t$ 
is a operator $\bar \Phi_t$ on ${\cal M}^{*}$, 
such that 
\[ (\bar \Phi_t (A) |B) = (A | \Phi_t (B)) \] 
for all $B \in D (\Phi_t) \subset {\cal M}$ 
and some $A \in {\cal M}^{*}$. 
The scalar product on ${\cal M}$
can be defined by
\[ (A|B)=\int dq dp \, [A(q,p)]^{*} B(q,p) \]
Then an operator $\bar \Phi_t$ is called adjoint if
\[ \int dq dp \, [(\bar \Phi_t A)(q,p)]^{*} B(q,p)=
\int dq dp \, [A(q,p)]^{*} (\Phi_t B)(q,p) . \]
Let us give the basic statement regarding the adjoint operator. 
If $\bar \Phi_t$ is an adjoint operator of $\Phi_t$, 
then the operator
\[ \bar \Phi^{(\alpha)}_t =
\int^{\infty}_0 ds f_{\alpha}(t,s) \bar \Phi_s , 
\quad (t>0) , \]
is an adjoint operator of $\Phi^{(\alpha)}_t$.
We prove this statement by using the Bochner-Phillips formula:
\[ \int dq dp \, (\bar \Phi^{(\alpha)}_t A)^{*} B =
\int^{\infty}_0 ds \, f_{\alpha}(t,s) 
\int dq dp \, (\bar \Phi_s A)^{*} B = \]
\[ =\int^{\infty}_0 ds \, f_{\alpha}(t,s) 
\int dq dp \, A^{*} (\Phi_s B) = 
\int dq dp \, A^{*} (\Phi^{(\alpha)}_t B) . \]
The semi-group $\{\bar \Phi_t, t>0\}$ describes a temporal evolution 
of the distribution function $\rho_t(q,p)=\bar \Phi_t \rho_0(q,p)$
by the Liouville equation
\[ \frac{d}{dt} \rho_t(q,p)=-\bar {\cal L} \rho_t(q,p) ,  \]
where
\[ \bar {\cal L}={\cal L}+ \Omega(q,p) , 
\quad \Omega(q,p)= \sum^n_{k=1} \Bigl( \frac{\partial G_k}{\partial q_k} +
\frac{\partial F_k}{\partial p_k}\Bigr) . \]
If $\Omega<0$, then the system is called dissipative.
The semi-group $\{\bar \Phi^{(\alpha)}_t, t>0\}$
describes the evolution of the density function
\[ \rho_t(\alpha,q,p)=\bar \Phi^{(\alpha)}_t \rho_0(q,p) \]
by the fractional equation
\[ \frac{d}{dt} \rho_t=-(\bar {\cal L})^{\alpha} \rho_t. \]
This is the {\it fractional Liouville equation}. \\

(4) It is known that $\bar \Phi_t$ is a real operator if
$\Phi_t$ is real. 
Analogously, if $\Phi^{(\alpha)}_t$ is a real operator, then
$\bar \Phi^{(\alpha)}_t$ is real. \\

(5) Let $\Phi_t$, $t>0$, be a positive one-parameter operator, i.e.,
\[ \Phi_t A \ge 0  \]
for $A \ge 0$. 
Using the Bochner-Phillips formula and the property
\[ f_{\alpha}(t,s) \ge 0  \quad (s>0) , \]
it is easy to prove that
\[ \Phi^{(\alpha)}_t A \ge 0  \quad (A \ge 0), \]
i.e.,  the operators $\Phi^{(\alpha)}_t$
are also positive.


\section{Examples of fractional equations of motion}

\subsection{Fractional free motion of particle}

Let us consider equation (\ref{Ham1}) for free particle.
Then 
\[ H=\frac{1}{2m} p^2 , \quad {\cal L}=\{H, \ . \ \}=\frac{p}{m} \partial_q ,\]
where $p$ is dimensionless variable
and $m^{-1}$ has the action dimension.
For $A=q$, and $A=p$, equation (\ref{Ham1}) gives
\[ \frac{d}{dt} q_t=\frac{1}{m} p_t, \quad 
\frac{d}{dt} p_t=0 . \]
The well-known solutions of these equations are
\be \label{Hsol1}
q_t=q_0 +\frac{t}{m} p_0 , \quad p_t=p_0 . \ee
Using these solutions and the Bochner-Phillips formula, 
we can obtain solutions of the fractional equations
\be \label{ex1}
\frac{d}{dt} q_t=- \frac{1}{m^{\alpha}} \Bigl( p \partial_q\Bigr)^{\alpha} q_t, 
\quad \frac{d}{dt} p_t=0 . \ee
in the form
\[ q_t(\alpha)=\Phi^{(\alpha)}_t q_0=
\int^{\infty}_0 ds f_{\alpha}(t,s) q_s , 
\quad p_t(\alpha)=p_0 , \]
where $q_s$ is given by (\ref{Hsol1}).
Then
\[ q_t=q_0 +\frac{1}{m} b_{\alpha}(t) p_0 , \quad p_t=p_0 , \] 
where
\[ b_{\alpha}(t)= \int^{\infty}_0 ds f_{\alpha}(t,s) \, s . \]

If $\alpha=1/2$, then we have
\[ b_{1/2}(t)= \frac{t}{2 \sqrt{\pi}} \int^{\infty}_0 ds \, 
\frac{1}{\sqrt{s}} e^{-t^2/4s} = \frac{t^2}{2} . \]
Then
\be \label{free}
q_t=q_0 -\frac{t^2}{2m^2} p_0 , \quad p_t=p_0 . \ee
These equations describe a fractional free motion for $\alpha=1/2$.

\subsection{Fractional equation for harmonic oscillator}

Let us consider equation (\ref{Ham1}) for harmonic oscillator.
Then ${\cal L}=\{H, \ . \ \}$, where
\be \label{oscHam}
H=\frac{1}{2m} p^2 +\frac{m\omega^2}{2} q^2, \ee
where $t$ and $p$ are dimensionless variables.
For $A=q$, and $A=p$, equation (\ref{Ham1}) gives
\be \label{19}
\frac{d}{dt} q_t=\frac{1}{m} p_t, \quad 
\frac{d}{dt} p_t=-m \omega^2 q_t . \ee
The well-known solutions of these equations are
\[ q_t=q_0 \cos (\omega t) +\frac{p_0}{m \omega} \sin (\omega t) , \]
\be \label{osc1}
p_t=p_0 \cos (\omega t) - m \omega q_0 \sin (\omega t) . \ee 
Using (\ref{osc1}) and the Bochner-Phillips formula, 
we can obtain solutions of the fractional equations
\be \label{ex2}
\frac{d}{dt} q_t=- (\{H, \ . \ \})^{\alpha} q_t , \quad 
\frac{d}{dt} p_t=- (\{H, \ . \ \})^{\alpha} p_t , \ee
where $H$ is defined by (\ref{oscHam}).
It can be written in the form
\[ \frac{d}{dt} q_t=-\frac{1}{m^{\alpha}} \Bigl( m^2 \omega^2 q \partial_p
- p \partial_q \Bigr)^{\alpha} q_t\]
\[ \frac{d}{dt} p_t=-\frac{1}{m^{\alpha}} \Bigl(m^2 \omega^2 q \partial_p
- p \partial_q \Bigr)^{\alpha} p_t . \]
It is easy to see that these equations with $\alpha=1$ give Eqs. (\ref{19}).
The solutions of fractional equations (\ref{ex2}) have the forms
\[ q_t(\alpha)=\Phi^{(\alpha)}_t q_0=
\int^{\infty}_0 ds f_{\alpha}(t,s) q_s , \]
\be \label{osc2} 
p_t(\alpha)=\Phi^{(\alpha)}_t p_0=
\int^{\infty}_0 ds f_{\alpha}(t,s) p_s . \ee
Substitution of (\ref{osc1}) into (\ref{osc2}) gives
\be \label{Hsol2a}
q_t=q_0 C_{\alpha}(t) +\frac{p_0}{m \omega} S_{\alpha}(t) , \ee
\be \label{Hsol2b}
p_t=p_0 C_{\alpha}(t) - m \omega q_0 S_{\alpha}(t) , \ee
where
\[ C_{\alpha}(t)=\int^{\infty}_0 ds \, f_{\alpha}(t,s)\, \cos(\omega s) , \]
\[ S_{\alpha}(t)=\int^{\infty}_0 ds \, f_{\alpha}(t,s)\, \sin(\omega s) . \]
Equations (\ref{Hsol2a}) and (\ref{Hsol2b}) describe 
solutions of fractional equations (\ref{ex2})
for classical harmonic oscillator. 

If $\alpha=1/2$, then 
\[ C_{1/2}(t)=\frac{t}{2 \sqrt{\pi}} \int^{\infty}_0 ds \, 
\frac{\cos(\omega s)}{s^{3/2}} \, e^{-t^2/4s} , \]
\[ S_{1/2}(t)=\frac{t}{2 \sqrt{\pi}} \int^{\infty}_0 ds \, 
\frac{\sin(\omega s)}{s^{3/2}} \, e^{-t^2/4s} . \]
These functions can be presented through the Macdonald function
(see \cite{Prudnikov}, Sec. 2.5.37.1.) such that
\[ C_{1/2}(t)= \Bigl(\frac{\omega t^2}{4 \pi}\Bigr)^{1/4} 
\Bigl[ e^{+ \pi i/8} K_{-1/2} \Bigl(2 e^{+\pi i/4} 
\sqrt{ \frac{\omega t^2}{4} } \Bigr)+
e^{-\pi i/8} K_{-1/2} \Bigl(2 e^{-\pi i/4} 
\sqrt{ \frac{\omega t^2}{4} } \Bigr) \Bigr] , \]
\[ S_{1/2}(t)= i \Bigl(\frac{\omega t^2}{4 \pi}\Bigr)^{1/4} 
\Bigl[ e^{+ \pi i/8} K_{-1/2} \Bigl(2 e^{+\pi i/4} 
\sqrt{ \frac{\omega t^2}{4} }\Bigr)-
e^{-\pi i/8} K_{-1/2} \Bigl(2 e^{-\pi i/4} 
\sqrt{ \frac{\omega t^2}{4} } \Bigr) \Bigr] , \]
where $\omega>0$, and $K_{\alpha}(z)$ is the Macdonald function 
\cite{SKM}, which is also called the modified Bessel functions 
of the third kind.

Note that fractional oscillators are objects of numerous investigations 
(see, for example, \cite{M,ZSE,Stanislavsky1,Stanislavsky2,Tof,RR})
because of different applications.

\subsection{Fractional equation for damped oscillator}

Let us consider oscillator with linear friction
\be \label{damped}
\frac{d}{dt} q_t=\frac{1}{m} p_t , \quad 
\frac{d}{dt} p_t=-m \omega^2 q_t-2 \beta p_t , \ee
where $\beta< \omega$. The solution of (\ref{damped}) has the form
\[ q_t= e^{-\beta t} \Bigl[ q_0 \cos (\sqrt{\omega^2-\beta^2} t) +
\frac{1}{m \omega} p_0 \sin (\sqrt{\omega^2-\beta^2} t) \Bigr], \]
\be \label{dos1}
p_t=e^{-\beta t}\Bigl[ p_0 \cos (\sqrt{\omega^2-\beta^2} t) - 
m \omega q_0 \sin (\sqrt{\omega^2-\beta^2} t) \Bigr] . \ee 
The fractional equations has the form
\[ \frac{d}{dt} q_t=- \Bigl((m \omega^2 q + 2 \beta p)\partial_p
- \frac{p}{m} \partial_q \Bigr)^{\alpha} q_t , \]
\[ \frac{d}{dt} p_t= - \Bigl((m \omega^2 q + 2 \beta p)\partial_p
- \frac{p}{m} \partial_q \Bigr)^{\alpha} p_t . \]
It is easy to see that these equations with $\alpha=1$ give Eqs. (\ref{damped}). 
Using (\ref{dos1}) and the Bochner-Phillips formula, 
we obtain the solutions 
\[ q_t=q_0 C_{\alpha, \beta}(t) +\frac{1}{m \omega} p_0 S_{\alpha, \beta}(t) , \]
\be 
p_t=p_0 C_{\alpha, \beta}(t) - m \omega q_0 S_{\alpha, \beta}(t) , \ee
where
\[ C_{\alpha, \beta}(t)=\int^{\infty}_0 ds \, f_{\alpha}(t,s)\, e^{-\beta t}
\cos(\sqrt{\omega^2-\beta^2} s) , \]
\be S_{\alpha, \beta}(t)=\int^{\infty}_0 ds \, f_{\alpha}(t,s)\, e^{-\beta t}
\sin(\sqrt{\omega^2-\beta^2} s) . \ee
These equations describe solutions of the fractional damped motion
of classical harmonic oscillator.


\end{document}